\documentclass[10pt,letterpaper]{article}
\usepackage{opex3}
\usepackage{subfigure}
\usepackage[doublespacing]{setspace}
\usepackage{ae} %%for Computer Modern fonts
\usepackage{gensymb}
\usepackage[normalem]{ulem}

\begin{document}

%%%%%%%%%%%%%%%%%% title page information %%%%%%%%%%%%%%%%%%
\title{High confinement, high yield Si$_3$N$_4$ waveguides for nonlinear optical applications}

\author{J\"{o}rn P. Epping,$^{1}$ Marcel Hoekman,$^{2}$ Richard Mateman,$^{2}$ Arne Leinse,$^{2}$ Ren\'{e} G. Heideman,$^{2}$ Albert van Rees,$^{3}$ Peter J.M. van der Slot,$^{1}$ Chris J. Lee,$^{1,4}$ and Klaus-J. Boller$^{1,*}$}

\address{$^{1}$Laser Physics \& Nonlinear Optics Group, Faculty of Science and Technology, MESA+ Institute for Nanotechnology, University of Twente, P. O. Box 217, Enschede 7500 AE, The Netherlands
 			\\$^{2}$LioniX BV, PO Box 456, Enschede 7500 AL, The Netherlands
 			\\$^{3}$XiO Photonics BV, PO Box 1254, Enschede 7500 BG, The Netherlands
 			\\$^{4}$XUV Optics Group, Faculty of Science and Technology, MESA+ Institute for Nanotechnology, University of Twente, P. O. Box 217, Enschede 7500 AE, The Netherlands}

\email{$^{*}$K.J.Boller@utwente.nl} %% email address is required

% \homepage{http:...} %% author's URL, if desired

%%%%%%%%%%%%%%%%%%% abstract and OCIS codes %%%%%%%%%%%%%%%%
%% [use \begin{abstract*}...\end{abstract*} if exempt from copyright]

\begin{abstract}
In this paper we present a novel fabrication technique for silicon nitride (Si$_{3}$N$_{4}$) waveguides with a thickness of up to 900~nm, which are suitable for nonlinear optical applications. The fabrication method is based on etching trenches in thermally oxidized silicon and filling the trenches with Si$_{3}$N$_{4}$. Using this technique no stress-induced cracks in the Si$_{3}$N$_{4}$ layer were observed resulting in a high yield of devices on the wafer. The propagation losses of the obtained waveguides were measured to be as low as 0.4~dB/cm at a wavelength of around 1550~nm.
\end{abstract}

%%%%%%%%%%%%%%%%%%%%%%% References %%%%%%%%%%%%%%%%%%%%%%%%%

%%%%%%%%%%%%%%%%%%%%%%%%%%  body  %%%%%%%%%%%%%%%%%%%%%%%%%%

\section{Introduction}

Silicon nitride-based waveguides form a promising, CMOS-compatible platform in integrated photonics research~\cite{Moss2013}. Especially, stoichiometric silicon nitride (Si$_3$N$_4$) deposited using low-pressure chemical vapor deposition (LPCVD) offers extremely low intrinsic losses and superior reproducibility. Further advantages of this platform are a broad transparency ranging from the visible~\cite{Romero-Garcia2013} to the mid-infrared, a high index contrast and absence of two-photon absorption in the near-infrared, including all the telecommunication bands. Due to their low propagation losses in the C-band as well as their low coupling losses by using spot-size converters Si$_3$N$_4$ waveguides are favored in the field of microwave photonics~\cite{Roeloffzen2013, Marpaung2013} and for novel types (glass-semiconductor) lasers with record low spectral bandwidth~\cite{Oldenbeuving2013,Fan2014}. Furthermore, ultra-low propagation losses in the C-band of 0.32~dB/m have been demonstrated with weakly confined modes using 40~nm thin Si$_3$N$_4$ waveguides~\cite{Spencer2014,Bauters2011}. The low confinement of the mode, however, has the drawback of increased bending losses and, as a consequence, limit the density of devices in integrated circuits. Further applications of Si$_3$N$_4$ waveguides are, e.g., bio-chemical applications such as optical coherence tomography~\cite{Yurtsever2014} and lab-on-a-chip devices due to the compatibility of Si$_3$N$_4$ waveguides with microfluidic channels~\cite{Cai2012,Ymeti2005}.

Furthermore, Si$_3$N$_4$ waveguides are of high interest for nonlinear integrated photonics~\cite{Moss2013} due to their high Kerr index~\cite{Ikeda2008}, while supporting highly confined modes due to their high index contrast and lacking nonlinear losses in the near-infrared. Nonlinear effects such as supercontinuum generation~\cite{Halir2012} and parametric frequency comb generation~\cite{Levy2009} have been demonstrated. For the latter, huge potential in nonlinear optical signal processing was shown by transmitting a data stream of 1.44~Tbit/s~\cite{Pfeifle2014}. It is desirable to select the waveguides such that the pump wavelength for broadband wavelength conversion lies in the spectral region where laser sources are readily available, i.e. 1000 to 2000~nm. However, to obtain phase matching for broadband wavelength conversion, the dispersion of the waveguide must be engineered such that pump wavelength is in the anomalous dispersion regime, while also being close to the zero dispersion wavelength (ZDW).

\begin{figure}[tbp]
	\centering
		\includegraphics[height=4.8cm]{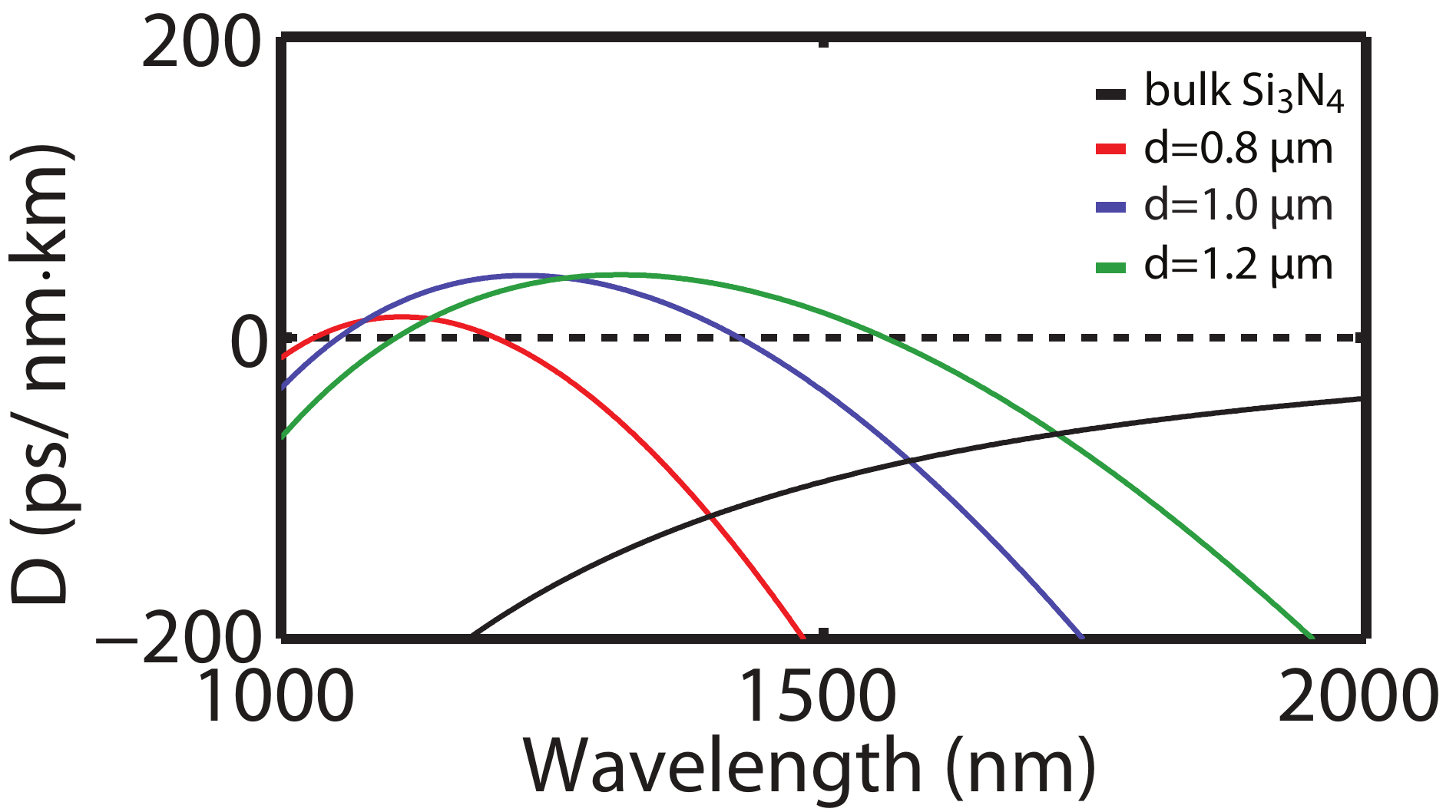}
	\caption{\label{fig:0} Calculated dispersion parameter, $D$, for bulk Si$_3$N$_4$ (black) and for quasi-TM modes (primarily polarized along the y-direction) of waveguides with a fixed width of 0.8~$\mu$m and a height of 0.8~$\mu$m (red), 1.0~$\mu$m (blue), and 1.2~$\mu$m (green). While bulk Si$_3$N$_4$ does not exhibit a ZDW at wavelengths between 1000~nm and 2000~nm, the ZDW of the waveguides are 1200~nm ($d=$ 0.8~$\mu$m), 1420~nm ($d=$ 1.0~$\mu$m), and 1560~nm ($d=$ 1.2~$\mu$m).}
\end{figure}

To achieve high modal confinement and to shift the ZDW to the near-infrared, the thickness of the silicon nitride waveguide core must be increased substantially from the values that are typically used ($< 200$~nm). To obtain the dispersion of the waveguides we simulated the effective refractive index $n$ for various wavelengths using a finite element solver (COMSOL Multiphysics) that takes both the dispersion of the materials and, as well, the dispersion induced by the geometry of the waveguide core into account. From this the dispersion parameter, $D = -\frac{\lambda}{c}\cdot\frac{d^{2}n}{d \lambda^{2}}$~\cite{Agrawal2007}, can be calculated, where $\lambda$ is the vacuum wavelength of the light and $c$ the speed of light. This is shown in Fig.\ref{fig:0}, where we plot $D$ as a function of the wavelength for waveguides with a fixed width $w = $ 0.8~$\mu$m and three different heights ranging from 0.8 to 1.2~$\mu$m. As a comparison, the dispersion parameter for bulk Si$_3$N$_4$ is shown as well. While bulk Si$_3$N$_4$ does not have a ZDW in the wavelength range from 1000 to 2000~nm, the ZDW for the waveguides is calculated to range from 1200~nm ($d=0.8$~$\mu$m, red) to 1560~nm ($d=1.2$~$\mu$m, green) by changing the thickness $d$ and, hence, providing phase matching for wavelengths around 1550~nm.

However, realizing Si$_3$N$_4$ waveguides with sufficient thickness is problematic. Typically, a Si$_3$N$_4$ layer is deposited on a thermally oxidized silicon substrate via LPCVD, and waveguide cores are etched from this Si$_3$N$_4$ layer. This layer has a limited thickness due to the increasing tensile stress that develops during the deposition of the Si$_3$N$_4$ with increasing layer thickness. Ultimately, this stress leads to the formation of cracks when the thickness is larger than about 400~nm~\cite{Luke2013}. Recently, improved methods for manufacturing low-loss thick Si$_3$N$_4$ waveguides have been introduced~\cite{Luke2013,Gondarenko2009}, however, they still suffer from cracks in the Si$_3$N$_4$ layer~\cite{Luke2013}, limiting the yield and reproducibility of device fabrication.

In this paper we report on the fabrication of the first silicon nitride waveguides with thicknesses greater than 800~nm in a reproducibly crack-free manner. Instead of depositing a thick layer of Si$_3$N$_4$, followed by etching the waveguide cores from this layer, we first etch trenches in the silicon dioxide cladding that is formed by thermally oxidizing the top surface of the silicon substrate. The trenches are then filled with Si$_3$N$_4$ using LPCVD and the resulting waveguide cores are crack-free for depths of up to $d = $ 1.2~$\mu$m. Note, that this method, in general, provides waveguides with an aspect ratio, $w/d > 1$, unlike commonly used Si$_3$N$_4$ waveguides. As a result, the width of the waveguide core is the critical parameter that controls stress-induced cracks, which is related to the thickness of the Si$_3$N$_4$ layers. Waveguides fabricated in this way are referred to as TriPleX$^{TM}$\cite{heideman2006,triplex}. Details of this process are given in section 2. We then discuss in section 3 the characterization of several thick waveguides that are manufactured according to this novel procedure and end with a discussion and conclusion in section~4.

\section{Fabrication}

\begin{figure}[tbp]
	\centering
	\mbox{
		\subfigure{
			\includegraphics[height=2.8cm]{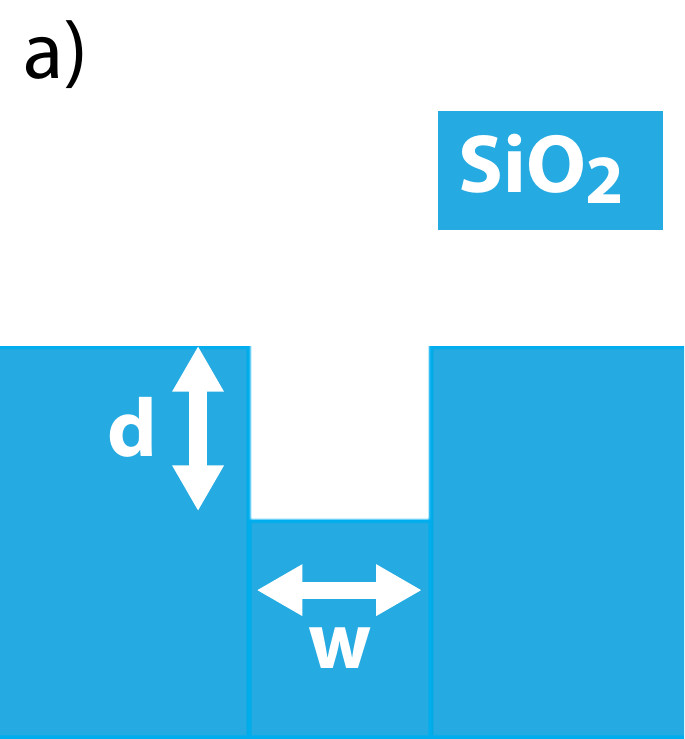}}
				\qquad
		\subfigure{
			\includegraphics[height=2.8cm]{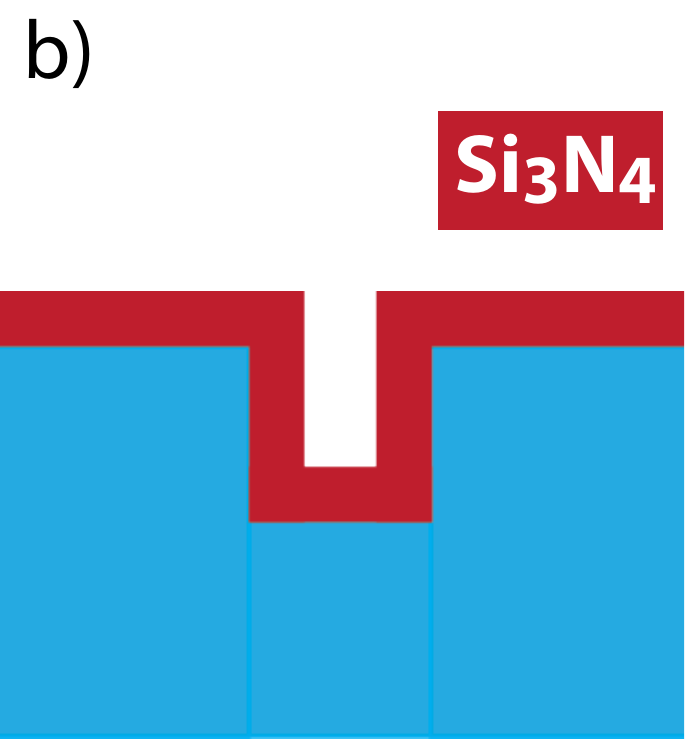}}
				\qquad
		\subfigure{
			\includegraphics[height=2.8cm]{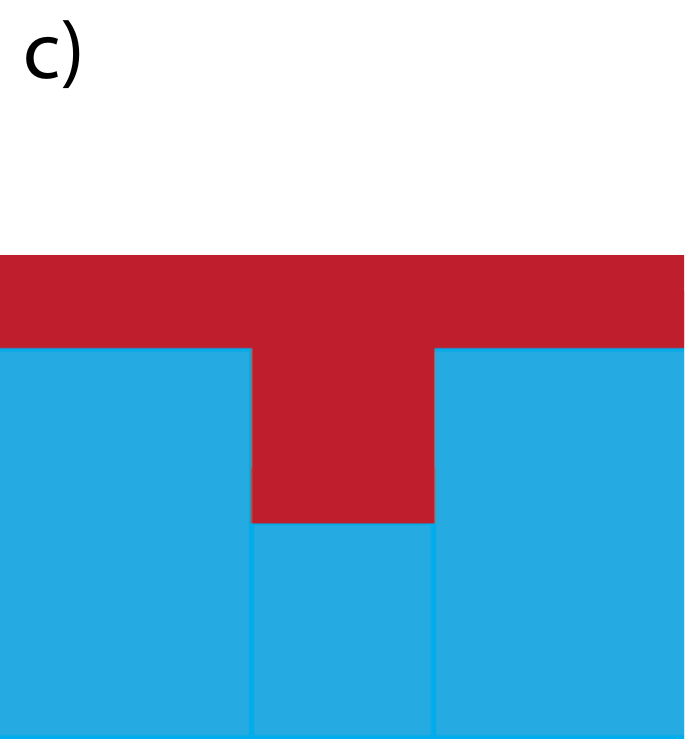}}
			\qquad
			\subfigure{
			\includegraphics[height=2.8cm]{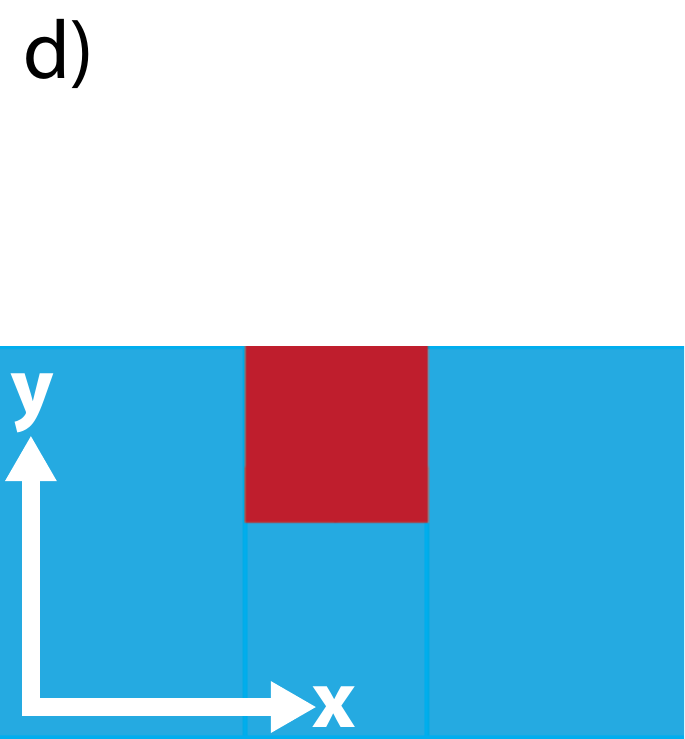}}
	}
	\caption{\label{fig:2} Steps of waveguide manufacturing. (a): A trench with depth $d$ and width $w$ is etched into an 8~$\mu$m layer of thermally oxidized silicon (SiO$_{2}$, blue) on a silicon wafer. (b): 250~nm of Si$_{3}$N$_{4}$ (red) is deposited homogeneously over all exposed surfaces using LPCVD. (c): The trench is completely filled with Si$_{3}$N$_{4}$ and a homogeneous Si$_{3}$N$_{4}$ layer over the whole trench width of up to 0.9~$\mu$m is formed. (d): The residual Si$_{3}$N$_{4}$ on the top surface of the wafer is removed from the wafer using chemical-assisted mechanical polishing and dry etching. The remaining Si$_{3}$N$_{4}$ forms the waveguide core with the dimensions of the etched trench. A top cladding of SiO$_2$ can be deposited optionally on top of the waveguide cores.}
\end{figure}

In this method, the waveguide devices are manufactured by filling predefined trenches with Si$_{3}$N$_{4}$ using LPCVD. The individual fabrication steps are shown in Fig.~\ref{fig:2}. In the first step, as shown in Fig.~\ref{fig:2}(a), reactive ion etching (RIE) is used to etch trenches with a width $w$ and a depth $d$ into the thermally oxidized top surface (oxide depth 8~$\mu$m) of a silicon wafer.

In the next step, a 250~nm thick layer of stoichiometric Si$_{3}$N$_{4}$ is deposited. During this deposition process, the Si$_{3}$N$_{4}$ layer grows equally thick from the sidewalls and bottom of the trenches, as illustrated in Fig.~\ref{fig:2}(b). To release stress from the material, the deposition is performed in steps and wafers are allowed to cool down to room temperature between depositions~\cite{Gondarenko2009}. As the deposition continues, the trench fills with Si$_{3}$N$_{4}$ until the the two layers at the side walls merge in the middle of the trench, such that the trench is completely filled, as shown in Fig.~\ref{fig:2}(c). For the manufactured waveguides no more than two deposition steps were required. The final waveguide core is obtained by removing the remaining Si$_{3}$N$_{4}$ from the top of the wafer using chemical-assisted mechanical polishing and dry etching, as shown in Fig.~\ref{fig:2}(d). To ensure that etching is stopped when the Si$_{3}$N$_{4}$ from the top of the wafer is removed an etch rate monitor is used. If required, this last processing step can also be applied between deposition steps, to thin the Si$_3$N$_4$ layer on top of the wafer to prevent cracks. To complete the waveguide cladding, first a layer of tetraethyl orthosilicate (Si(OC$_{2}$H$_{5}$)$_{4}$, TEOS) is deposited by LPCVD, which will form SiO$_2$ after annealing. Second, a layer of SiO$_2$ is grown to a thickness of 8~$\mu$m by plasma-enhanced chemical vapor deposition (PECVD) (both not shown in Fig.~\ref{fig:2}). To convert the TEOS layer into SiO$_2$ and to reduce the absorption in the near infra-red caused by dangling Si-H bonds, the wafers are annealed after deposition at 1150~$^{\circ}$C.

\begin{figure}[htbp]
	\centering
	\mbox{
		\subfigure{
			\includegraphics[height=2.8cm]{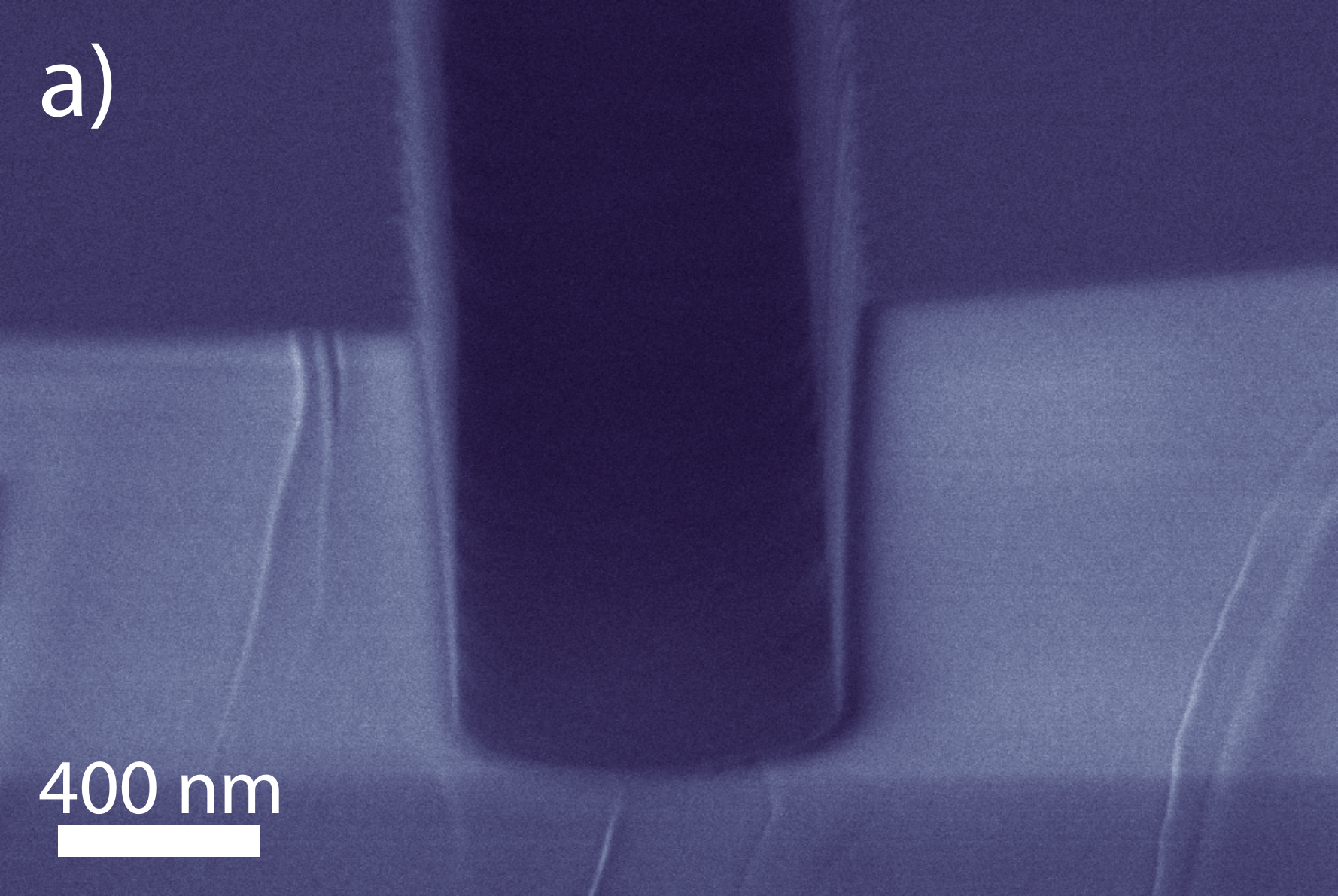}}
		\subfigure{
			\includegraphics[height=2.8cm]{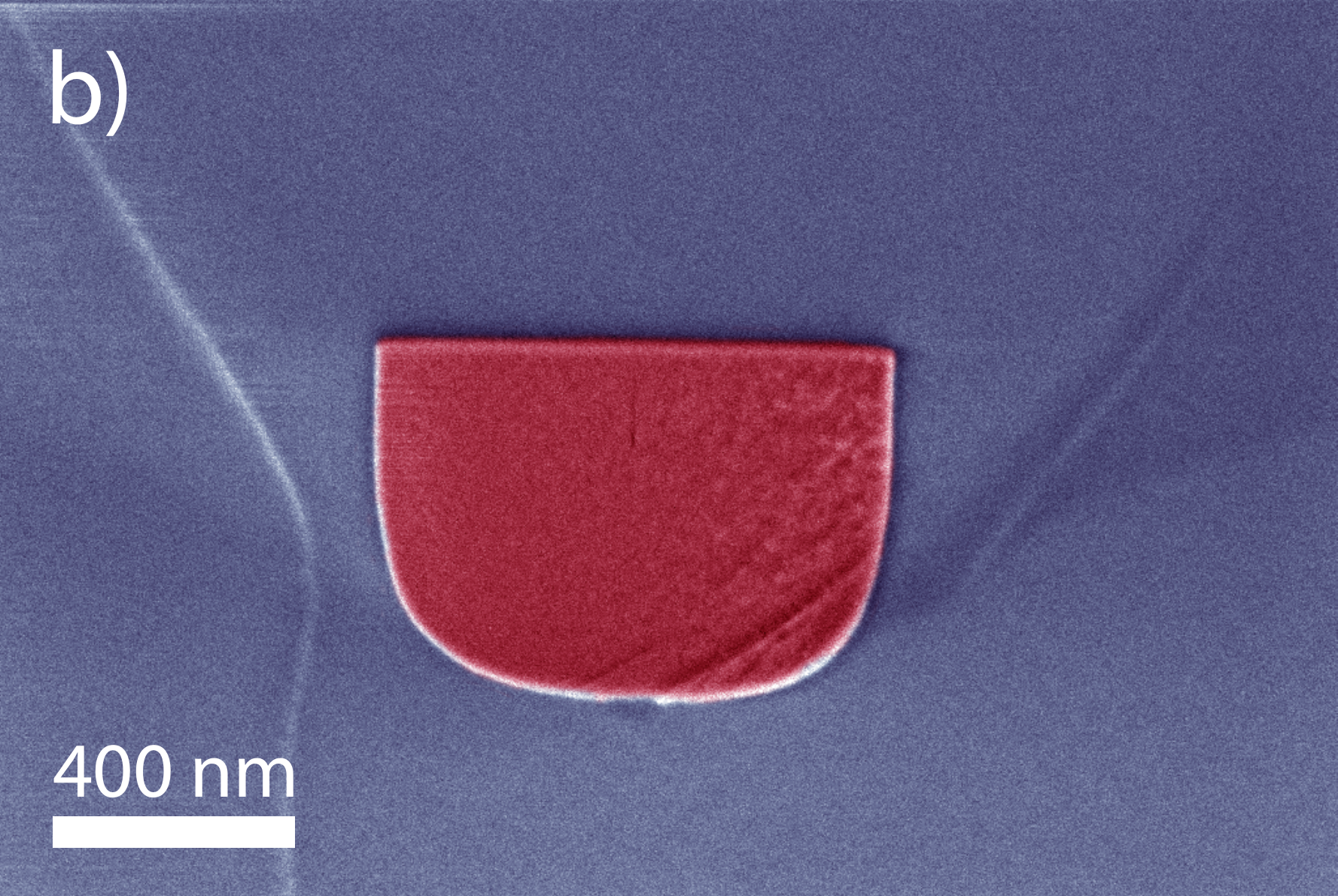}}
		\newline
		\subfigure{
			\includegraphics[height=2.8cm]{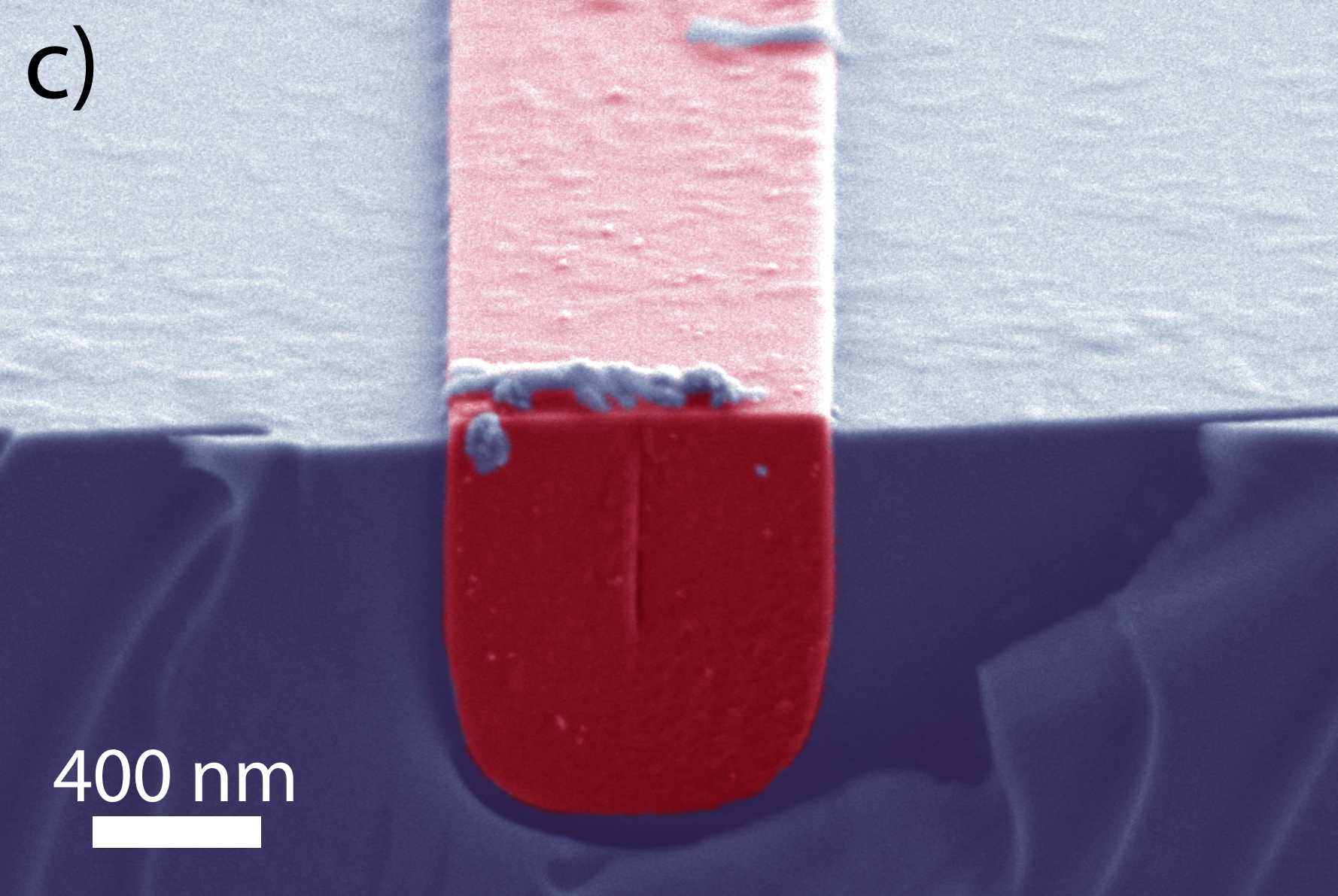}}
	}
	\caption{	\label{fig:1} SEM pictures: (a) A trench etched in thermally oxidized silicon with $w = 0.8$~$\mu$m and $d = 1.0$~$\mu$m. (b) and (c): the cross section of silicon nitride waveguide core with $w = 0.83$~$\mu$m and $d = 0.8$~$\mu$m with a layer PECVD silica on top (b) and $w = 0.9$~$\mu$m and $d = 1.2$~$\mu$m without a top layer (c). Note that all pictures are taken at an angle, making the vertical direction appear smaller.}
\end{figure}

Using this procedure, several TriPleX-based Si$_{3}$N$_{4}$ waveguides were manufactured with their width, $w$, ranging from 0.6 to 0.9~$\mu$m, and depths, $d$, of 0.8, 1.0, and 1.2~$\mu$m. To ensure that these waveguides have low propagation losses, it is important that the trenches have a low surface roughness at the bottom as well as at the side walls, since rougher surfaces will result in higher scattering losses for propagating light. A scanning electron microscope (SEM) picture of an etched trench is shown in Fig.~\ref{fig:1}(a) for a width of $w=$0.8~$\mu$m and a depth of $d=$1.0~$\mu$m. Note that the vertical direction of all the SEM pictures in Fig.~\ref{fig:1} appears to be smaller because the pictures were taken under an angle. Figure~\ref{fig:1}(a) also shows a slight positive taper, meaning that the width of the trench increases from bottom to top, which ensures a conformal deposition. Without tapering, the deposition is not fully conformal, and a gap between the silicon nitride and the trench wall appears, which increases the propagation losses as a consequence.

SEM pictures of two completed waveguide cores are shown in Fig.~\ref{fig:1}(b) and \ref{fig:1}(c) with a cross section of $w=$~0.83~$\mu$m by $d=$~0.8~$\mu$m and $w=$~0.9~$\mu$m by $d=$~1.2~$\mu$m, respectively, with and without a top cladding, respectively. To increase the contrast between the Si$_{3}$N$_{4}$ and the SiO$_2$ in the SEM pictures, the samples were etched, which resulted in the Si$_{3}$N$_{4}$ waveguide core sticking out from the SiO$_2$ because of a difference in etching rates of the materials. In Fig.~\ref{fig:1}(b) it can be seen that a homogeneous waveguide core with $w = 0.83$~$\mu$m and $d = 0.8$~$\mu$m is formed using this manufacturing method.

By comparing the shape of the trench as shown in Fig.~\ref{fig:1}(a) and the waveguides in Figs.~\ref{fig:1}(b) and \ref{fig:1}(c), the stress induced by the deposition of silicon nitride is evident. The large difference in material stress between the SiO$_{2}$ and Si$_{3}$N$_{4}$ layers leads to a change in both the shape of the waveguide as well as an inwards bend of the side walls. The rounded shape of the waveguides leads to a small ($\sim$10~nm) blue-shift in the ZDW compared to the rectangular shape assumed in Fig.\ref{fig:0}. However, the variations in waveguide dimensions, given by fabrication tolerances, cause larger shifts. In the case of the deepest etching depth ($d = 1.2$~$\mu$m), it is observed that the change of the waveguide shape is even stronger. Here, the tapering of the trench was insufficient to compensate for the sidewall bending induced by the tensile stress of the Si$_{3}$N$_{4}$ layer. This effects the final closure of the trench when the deposited Si$_{3}$N$_{4}$, which grows from the two side walls, meet in the middle. As a result, the trench is not filled entirely with Si$_{3}$N$_{4}$ as can be seen in Fig.~\ref{fig:1}(c). In future, we will compensate for this effect by applying a stronger positive taper to the sidewall of the trenches.

During the fabrication, a large number of waveguides and waveguide circuits were obtained with the described technique covering a total of six wafers (100~mm diameter) and not a single crack was observed in the Si$_{3}$N$_{4}$ layers. This clearly indicates that, using our novel approach, crack-free Si$_{3}$N$_{4}$ waveguides with a width of up to 0.9~$\mu$m can be fabricated.

\section{Characterization}

The propagation losses are an essential characteristic for integrated waveguides, since they are critical for the design of devices such as micro-resonators. By measuring the transmission of a broadband light source through the waveguides, the propagation losses were determined. The broadband light from a superluminescent diode with a center wavelength of 1560~nm and a spectral bandwidth at full width at half maximum of 45~nm was used. The light was injected and collected using lensed fibers (2~$\mu$m spot size at 1550~nm), while the input polarization was controlled to excite the quasi-TM mode which is primarily polarized along the y-direction (see Fig.~\ref{fig:2}(d)) of the waveguides using an in-line optical fiber polarization controller. The transmitted power was measured for three etching depths, 0.8~$\mu$m, 1.0~$\mu$m, and 1.2~$\mu$m, and for four different propagation lengths, using spiral waveguides with lengths of 2.30~cm, 3.89~cm, 6.90~cm, and 10.48~cm and of various widths.

\begin{figure}[htbp]
	\centering
	\mbox{
		\subfigure{
				\includegraphics[height=5cm]{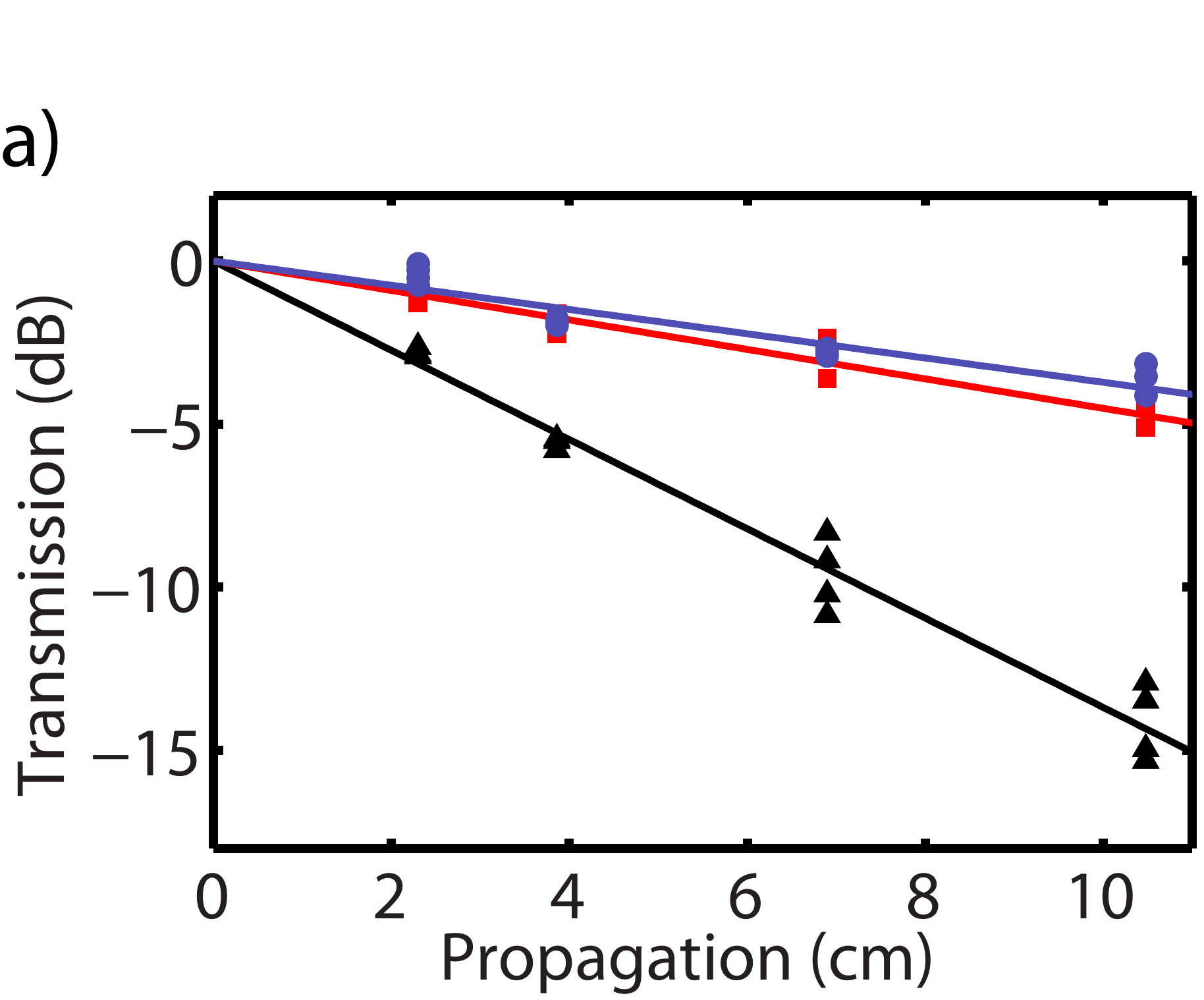}}
				\qquad
		\subfigure{
			\includegraphics[height=5cm]{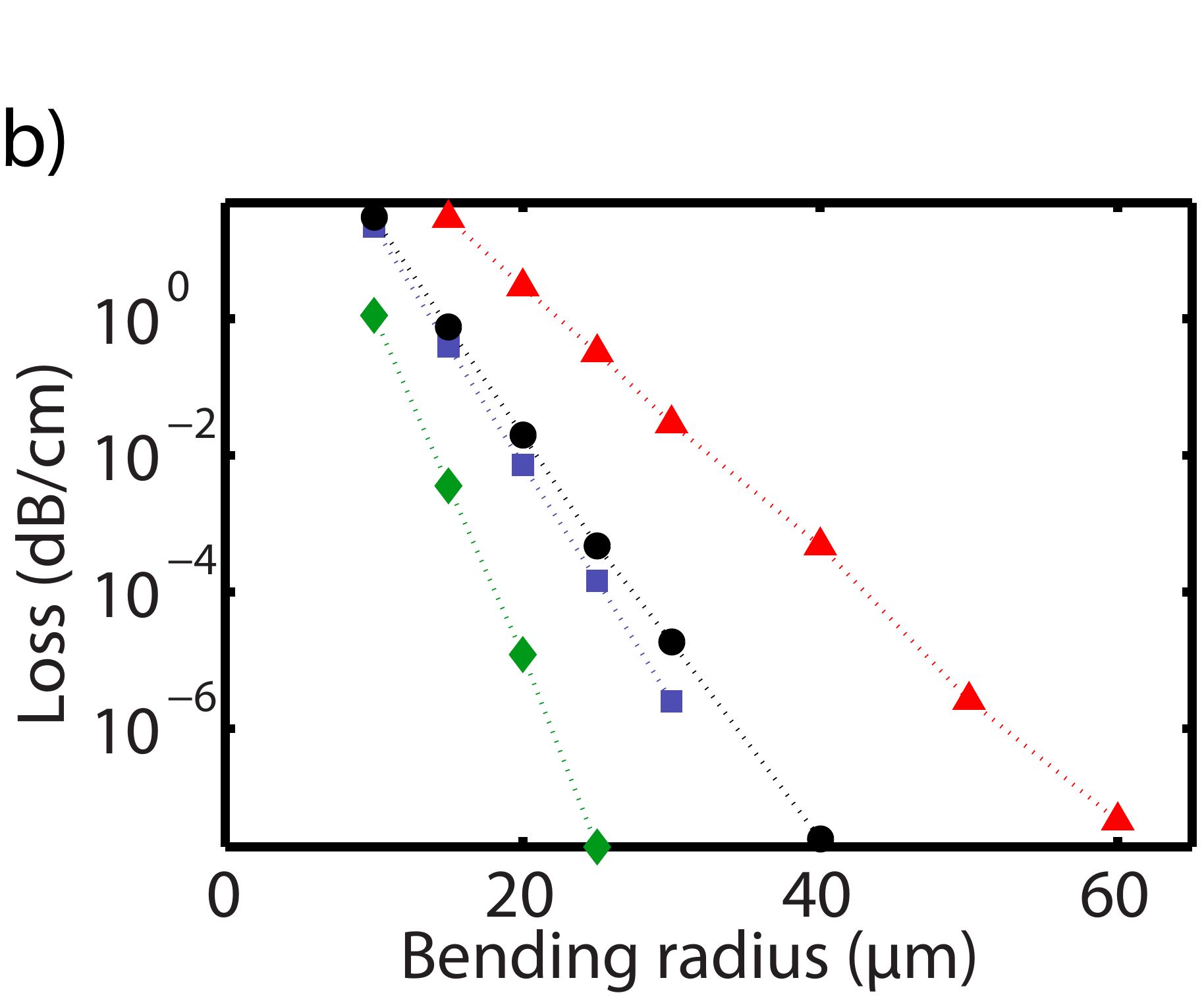}}
	}
	\caption{	\label{fig:3} (a) Power transmission of light with a center wavelength of 1560~nm (TM polarization) measured for different waveguide lengths (2.30~cm, 3.89~cm, 6.90~cm, and 10.48~cm) and waveguide depths (0.8~$\mu$m (blue circles), 1.0~$\mu$m (red squares), and 1.2~$\mu$m (black triangles)). The waveguide width varies between 0.7~$\mu$m and 0.9~$\mu$m. The graph has been corrected for the coupling losses, which vary between 13 to 15~dB in total. (b) Simulated bending losses of the quasi-TM mode at a wavelength of 1550~nm for various bending radii. Simulations were performed for cross sections and depths of $w=0.5$~$\mu$m and $d=0.8$~$\mu$m (red triangles), $w=0.8$~$\mu$m and $d=0.8$~$\mu$m (blue squares), $w=0.5$~$\mu$m and $d=1.2$~$\mu$m (black circles), and $w=0.8$~$\mu$m and $d=1.2$~$\mu$m (green diamonds).}
\end{figure}

The transmission as a function of waveguide length is shown in Fig.~\ref{fig:3}(a), and from this data, the propagation loss coefficient is determined to be 0.37~dB/cm ($d=0.8$~$\mu$m, blue), 0.45~dB/cm ($d=1.0$~$\mu$m, red), and 1.37~dB/cm ($d=1.2$~$\mu$m, black), while $w$ ranged from 0.7 to 0.9~$\mu$m. No variation in the propagation loss was observed for the range of waveguide widths investigated. The propagation losses are believed to be mainly due to scattering losses since absorptive losses have been reported to amount to values of below 0.055~dB/cm in Si$_{3}$N$_{4}$ waveguides~\cite{Gondarenko2009}. We address the comparably high losses for a $d$ of 1.2~$\mu$m to higher scattering losses resulting from the small gap in the center of the Si$_{3}$N$_{4}$ core, as shown in Fig.~\ref{fig:1}(c).

Another important performance parameter is the bending loss, as this indicates how densely devices can be packed on a wafer. In our measurements we found that the bending losses are so low that we could not experimentally distinguish these from the differences in coupling efficiencies from waveguide to waveguide. To quantify bending losses for thick Si$_{3}$N$_{4}$ waveguides, we simulated them using a finite element solver (FieldDesigner, PhoeniX BV). The simulated bending losses for a quasi-TM mode at a wavelength of 1550~nm are shown in Fig.~\ref{fig:3}(b) as a function of the radius of the bend for waveguide cross sections of $d = 0.8$~$\mu$m and 1.2~$\mu$m and $w = 0.5$~$\mu$m and 0.8~$\mu$m, respectively. As a result of the high confinement, the bending losses of the guided mode are lower for larger waveguide cross sections. As shown in Fig.~\ref{fig:3}(b), e.g., the losses for a bending radius of 20~$\mu$m decrease from 2.9~dB/cm for the smallest waveguide cross section, $w=0.5$~$\mu$m and $d=0.8$~$\mu$m (red), to as low as $1.2 \cdot 10^{-5}$~dB/cm for $w=0.8$~$\mu$m and $d=1.2$~$\mu$m (green). The bending losses are negligible for the bigger cross sections when compared to the propagation losses. The simulated bending loss for waveguides with high modal confinement and $d = w = 0.8$~$\mu$m (blue) is only 0.7~dB/m for a bending radius of 20~$\mu$m. In comparison, to obtain a similar bending loss from a waveguide with low modal confinement~\cite{Bauters2011}, the bending radius must be as large as 2~mm or more. Consequently, the density of waveguides circuits on a wafer can be increased by orders of magnitude using waveguides with a high confinement, which leads to a superior efficiency in the fabrication of Si$_{3}$N$_{4}$ waveguides.

\section{Conclusion}
In conclusion, we have shown a novel way to fabricate TriPleX-based Si$_{3}$N$_{4}$ waveguides with a width of up 0.9~$\mu$m, while having a depth of up to 1.2~$\mu$m. This method is promising to achieve a high yield compared to previously reported approaches to fabricate Si$_{3}$N$_{4}$ waveguides of similar cross-section dimensions. The complete area of six wafers with a diameter of 100~mm was used to produce waveguide structures, while not a single stress-induced crack appeared during or after fabrication. The propagation losses are measured to lie below 0.4~dB/cm for waveguides with a depth of 0.8~$\mu$m, while the bending losses were simulated to be below 0.01~dB/cm for a bending radius of 20~$\mu$m, both for a quasi-TM mode at a wavelength of 1550~nm. 

The measured coupling efficiency and observed variation in coupling efficiency can be improved by implementing spot-size converters at either end of the waveguide. The spot size converter expands the mode field diameter, which makes the coupling easier and less sensitive for imperfections in the flatness of the end facet. 

\section*{Acknowledgments}
This research is supported by the Dutch Technology Foundation STW, which is part of the Netherlands Organisation for Scientific Research (NWO), and which is partly funded by the Ministry of Economic Affairs

\end{document}